# Probing low-energy hyperbolic polaritons in van der Waals crystals with an electron microscope

*Alexander A. Govyadinov[1,*], Andrea Konečná[2,*], Andrey Chuvilin[1,3], Saül Vélez[1], Irene Dolado[1], Alexey Y. Nikitin[1,3], Sergei Lopatin[4], Fèlix Casanova[1,3], Luis E. Hueso[1,3], Javier Aizpurua[2,5], Rainer Hillenbrand[1,3,6,+]*

[1]CIC nanoGUNE, 20018 Donostia-San Sebastián, Spain
[2]Material Physics Center, CSIC-UPV/EHU, 20018 Donostia-San Sebastián, Spain
[3]IKERBASQUE, Basque Foundation for Science, 48011 Bilbao, Spain
[4] King Abdulla University of Science & Technology, Thuwal, 23955, Saudi Arabia
[5]Donostia International Physics Center DIPC, 20018 Donostia-San Sebastián, Spain
[6]UPV/EHU, 20018 Donostia-San Sebastián, Spain
[*]both authors contributed equally to this work.
[+]Corresponding Author: r.hillenbrand@nanogune.eu

**ABSTRACT**

Van der Waals (vdW) materials exhibit intriguing structural, electronic and photonic properties. Electron Energy Loss Spectroscopy (EELS) within Scanning Transmission Electron Microscope (STEM) allows for nanoscale mapping of such properties. However, typical STEM-EELS detection is limited to energy losses in the eV range, which are too large for probing important low-energy excitation such as phonons or mid-IR plasmons. Here we adapt a conventional STEM-EELS system to probe energy loss down to 100 meV, and apply it to map phononic states in h-BN, a representative van der Waals (vdW) material. The h-BN EELS spectra, surprisingly, depend on the h-BN thickness and the distance of the electron beam to h-BN flake edges. To explain this observation, we developed a classical response theory describing the interaction of fast electrons with (anisotropic) slabs of vdW materials. Theoretical and numerical calculations perfectly match the experimental h-BN spectra, and reveal that the electron loss is dominated by the excitation of hyperbolic phonon polaritons, and not of bulk phonons as often reported. Thus, STEM-EELS offers the possibility to probe low-energy polaritons in vdW materials, with our theory being of fundamental importance for interpreting future experiments.

Van der Waals (vdW) materials are a large class of layered materials arranged from individual atomic layers that are bound by (weak) van der Waals interactions[1]. Simple stacking of layers with different physical properties allows for creation of efficient, compact heterostructures that offer unique possibilities for development of electronic, photonic and optomechanical devices[1–6]. Much of the vdW materials functionality results from the large anisotropy in the bonding strength of atoms in the direction parallel to atomic layers and across them, and therefore is often intimately connected with the corresponding phonons. The investigation of phononic excitations in vdW materials thus requires probing low mid-IR energies at high spatial resolution (few tens of nm and below).

Electron energy loss spectroscopy performed in scanning transmission electron microscopy (STEM-EELS) allows for spatially-resolved mapping of a variety of excitations in condensed matter and can provide unprecedented structural information about the sample[7]. However, detecting low-energy excitations (<250 meV) with conventional STEM-EELS is currently more than a challenge. The poor monochromaticity of the (primary) electron beam and the limited resolution of the detection system yield a zero energy loss peak (ZLP) with a typical energy width of about 200 meV and larger[8]. The ZLP acts as a strong background, which prevents measurements of the low-energy loss signals from the sample.

Here, we decreased the ZLP width of a conventional STEM-EELS system to below 50 meV, which allowed us to probe low-energy phononic excitations (down to 100 meV) in vdW materials, particularly in the upper Reststrahlen band[9] (169-200 meV) of the hexagonal Boron Nitride (h-BN). We find energy loss peaks that cannot be interpreted as pure transverse (TO) or longitudinal (LO) optical phonon excitations[10]. We therefore develop a rigorous theory to understand the electron energy loss (EEL) in slabs of vdW materials (that show strong optical anisotropy) and demonstrate that it is dominated by the excitation of phonon polaritons, which exhibit a hyperbolic dispersion owing to the layered crystal structure of h-BN. We further show evidence that the electron beam excites hyperbolic surface phonon-polaritons[11,12] at the edges of h-BN flakes.

**RESULTS**

**STEM-EELS at low-energies.** For our experiments we used a FEI Titan 60-300 STEM-EELS system with 60 keV beam energy, in which we reduced the ZLP width from 200 meV (full width at half maximum, FWHM, according to the tool specifications) to 46 meV. This was achieved essentially by (i) enhancing the monochromaticity of the primary electron beam via the reduction of the anomalous energy broadening due to the Coulomb interaction of the beam electrons (Boersch effect[13]); and (ii) improving the energy resolution of the detection system by a five-fold increase of the spectrometer dispersion (see Methods for details). These modifications suppressed the spectral background (caused by ZLP tails) at energies down to about 100 meV and allowed us to probe excitations in the upper Reststrahlen band of h-BN.

We note that compared to recent developments (using cold field emission) that achieve better resolution[14], our development is based on a conventional system (using thermionic emission) and offers the advantage of higher and more stable currents[15,16], thus faster imaging.

**Spectral map of h-BN flake.** The sample studied in this work is an h-BN flake on a 15 nm thick $Si_3N_4$ membrane as shown in Fig. 1a. The black area corresponds to the bare membrane and the bright one to the h-BN flake (see inset for system topology). The brightness in this image is indicative of the h-BN thickness: the dark gray part corresponds to a few nanometers and the white part to about 30 nm, as estimated separately (after STEM-EELS) by atomic force microscopy of the same sample (see Methods for details).

Fig. 1b shows typical EEL spectra acquired on the bare membrane (magenta) and on the h-BN flake (open symbols). Both spectra are dominated by the ZLP (in which we include the energy loss in the Si$_3$N$_4$ membrane for simplicity) of nearly identical shape and FWHM of 46 meV, which demonstrates an excellent stability and consistency of our measurements at different spatial locations. A magnified view onto the h-BN spectrum (open symbols in Fig. 1c) reveals an energy loss peak in the range of 120 meV to 250 meV that is not present in the bare membrane spectrum (magenta), and thus indicates an excitation related to phonons in the h-BN. We can isolate this h-BN loss peak (blue dots in Fig. 1c) by subtracting the EELS spectrum of the membrane (see methods for details). Subsequent Gaussian fitting of the peak (see red curve in Fig. 1c and methods for details) yields its spectral position $\omega_0$. Isolating and fitting the peak at each pixel yields a spatial map of $\omega_0$, which is shown in Fig.1d. Intriguingly, we find that the peak position varies throughout the h-BN flake, revealing two notable trends. Firstly, $\omega_0$ depends on the h-BN film thickness gradually blueshifting from ~175 meV at the thinnest part (left flake extremity, see Fig. 1a to correlate with the flake thickness), to about 185 meV at the thickest part near the flake edge (blue zone at the bottom right). Secondly, the h-BN energy loss spectra for aloof beam positions[17] (i.e. outside the h-BN flake) are redshifted by up to 10 meV compared to the energy loss spectra of a beam passing through the h-BN flake (see bottom right part of the flake in Fig. 1d). Such surprising behavior cannot be explained by excitation of the TO or LO phonons, whose spectral positions are determined solely by the material properties and thus should not vary with thickness.

**Interaction of fast electrons with vdW materials.** To understand the variation of the spectral position of the energy loss peak in h-BN, we developed a theory for EEL in thin slabs of vdW materials, based on a classical electrodynamics approach to describe the electron-sample interaction. The electron is represented as a line current (non-recoil approximation), which induces an electromagnetic field, $\mathbf{E}_{\text{ind}}$, in the sample that acts back on the electron[18–20]. $\mathbf{E}_{\text{ind}}$ depends on the dielectric properties of the slab, which are described by the macroscopic permittivity $\epsilon$. Our approach rigorously accounts for the strong uniaxial anisotropy of h-BN by treating $\epsilon$ as a tensor[21] with two principal components $\epsilon_{\parallel}$ and $\epsilon_{\perp}$, which describe the dielectric response along and across the optical axis, respectively (the optical axis is perpendicular to the atomic layers of the vdW crystal, see Fig. 2). The probability $\Gamma$ of an electron to lose the energy $\hbar\omega$ can then be calculated as the Fourier transform of the work performed by the electron against this induced field along the entire electron trajectory, and (assuming cylindrical symmetry) can be written as (see Supplementary Information 1):

$\Gamma(\omega) = \int_0^{q_c} P(q,\omega)dq$, (1)

where $P(q,\omega) = \frac{e}{2\pi^3 \hbar \omega} \text{Re}\left[\int_{-\infty}^{\infty} dz\, \hat{z}\cdot \mathbf{E}_{\text{ind}}(q,z,\omega)\exp\left(\frac{-i\omega z}{v}\right)\right]$ is the probability for the electron to transfer a transverse (to the electron trajectory) momentum $q$ for the particular energy loss ($e$, $\hbar$ and $v$ are the electron charge, the Plank's constant and the relativistic speed of the electron, respectively). $q_c$~1 Å$^{-1}$ is the maximum transverse momentum of electrons that are allowed to pass through the collection aperture of the microscope detector. It is determined by a half-aperture collection angle of about 8 mrad of our STEM-EELS instrument. The integral in $P$ is taken over the electron trajectory (parallel to the optical axis z here) and can be decomposed into three contributions (see Supplementary Information 2):

$$P(q,\omega) = P_{\text{bulk}}(q,\omega) + P_{\text{guid}}(q,\omega) + P_{\text{begr}}(q,\omega) \qquad (2)$$

Here, $P_{\text{bulk}}$ corresponds to the energy loss due to excitations in bulk h-BN (i.e. modes of the unbound medium), similarly to that in isotropic samples[18,19]; $P_{\text{guid}}$ is related to the excitation of electromagnetic modes guided along the slab (analogous to the surface polariton waves in isotropic samples[22,23]) and finally, $P_{\text{begr}}$ accounts for the begrenzungseffekt, i.e. the reduction of the bulk loss due to energy transfer to the guided modes[7,20,24] (see schematics in Fig. 3a).

In Fig. 3c-e we show $P_{\text{bulk}}$, $P_{\text{guid}}$ and $P_{\text{begr}}$ as a function of $q$ and $\omega$, calculated for a 30 nm thick h-BN slab. We see that $P_{\text{bulk}}(q,\omega)$ (see equation S13) depicted in Fig. 3c is mainly due to the excitation of longitudinal phonon (see Supplemental Note 4) and thus gives rise to a strong bulk loss peak at the LO phonon frequency in $\Gamma_{\text{bulk}}(\omega)$ (blue curve in Fig. 3f). In contrast, $P_{\text{guid}}$ (Fig. 3d) strongly contributes to the energy loss in the lower half of the Reststrahlen band, yielding a broad peak in $\Gamma_{\text{guid}}(\omega)$ (red curve in Fig. 3f). $P_{\text{guid}}$ (equation S11) stems from hyperbolic phonon polaritons (HPhPs)[25,26] propagating through the volume of strongly anisotropic materials when $\epsilon_{\parallel}\epsilon_{\perp} < 0$. The later condition is satisfied in the upper Reststrahlen band of h-BN (Fig. 2), resulting in a dispersion that is described by a hyperboloid in the momentum space[27]. Upon excitation by a fast electron, HPhPs propagate in the h-BN volume as rays[5,25,28,29]. In a thin slab, the rays reflect at the top and bottom surfaces, forming a zig-zag pattern (see right hand side of the schematics in Fig. 3a), which can be described as a superposition of a number of guided HPhP modes M0, M1, M2,...[29] (the intensity profiles of the M0 to M2 are sketched in Fig. 3b). Each mode has a unique dispersion that relates its energy and momentum which can be visualized by plotting the imaginary part of the reflection coefficient $R_p$ of the slab[30]. For thin slabs of highly reflective materials (such as h-BN), $P_{\text{guid}}$ becomes proportional to $\text{Im}(R_p)$ (see Supplementary Information 3):

$$P_{\text{guid}}(q,\omega) \approx F(q,\omega)\text{Im}[q^{-1}R_p(q,\omega)], \qquad (3)$$

where $F(q,\omega) = \frac{2e}{\epsilon_0}\frac{q^3 v^2}{(q^2 v^2+\omega^2)^2}$ ($\epsilon_0$ is the permittivity of vacuum). The dominant contribution to $P_{\text{guid}}$ comes from the M0-HPhP mode. This can be verified by plotting its dispersion curve in Fig. 3d (dashed blue curve), which perfectly traces $P_{\text{guid}}(q,\omega)$.

$P_{\text{begr}}(q,\omega)$ (see equation S14) is depicted in Fig. 3e. This contribution is positive around the dispersion curve of the M0-HPhP mode (dashed blue line in Fig. 3e), but becomes negative at $(q,\omega)$-values where the bulk loss $P_{\text{bulk}}(q,\omega)$ assumes its maximum values. Consequently, $\Gamma_{\text{begr}}(\omega)$ exhibits a strong negative peak at the LO energy (gray curve in Fig. 3f), which is nearly identical in shape and strength to $\Gamma_{\text{bulk}}(\omega)$ (blue curve in Fig. 3f).

Importantly, upon summation of all three terms in equation (2), the begrenzungseffekt nearly completely cancels the bulk energy loss (as in EELS of isotropic materials[31], see equation S30), leaving only a small shoulder around the LO energy (green curve in Fig. 3f). Therefore, the EEL spectrum of thin vdW slabs is predominantly caused by the excitation of hyperbolic polaritons (HP), with the associated spectral peak located neither at TO nor LO frequencies. This central finding of our analysis has been often overlooked, resulting in incorrect assignment of phonon energies from EELS spectra[14].

In order to understand the position of the main electron energy loss peak associated with the HPhP excitation (marked by dashed vertical arrow in Fig 3f), we revisit equation (3). In this equation, $F$ can be understood as the "probability" for the electron (upon loosing energy $\hbar\omega$) to excite the dominant M0-HPhP mode at momentum $q$. This probability is the highest when $q = \sqrt{3}\omega/v$ (purple line in the inset of Fig. 3d, referred to as "electron line" in the following), yielding the momentum for the strongest excitation of the guided mode. The frequency $\omega_g$ at which this occurs determines the position of the spectral maximum in $\Gamma_{\text{guid}}(\omega)$ (see Supplemental Information 3 for details) and can be found geometrically as the intersection (marked by the purple dashed arrow in the inset of Fig. 3d, inset) of the "electron line" with the dominant M0-HPhP dispersion curve (dashed blue curve). This method of finding the peak position also applies for polariton excitations in isotropic materials[19] (see Fig. S2), making it a valuable universal tool for analyzing and interpreting the loss peaks in STEM-EELS.

**Thickness dependence**. Using our theoretical framework, we can explain the thickness dependence of the energy loss spectra of h-BN samples. To this end, we calculate EEL spectra for 2 nm, 15 nm and 30 nm thick h-BN slabs according to equation (1). For all spectra (Fig. 4b) we observe a dominant loss peak due to excitation of M0-HPhP mode. Most importantly, this peak shifts to lower energies with decreasing slab thickness, which is in good agreement with our experimental observations (Fig. 1d). This peak shift can be explained by the increasing momentum q of the M0-HPhP mode with decreasing thickness of the h-BN slab[30,32]. As can be seen in Fig. 4a, the intersection of the M0-HPhP mode dispersion (blue, red, green curves) with the "electron line" – and thus the maximum of the loss peak – shifts to lower energies.

For a quantitative comparison between experiment and theory we convolve the calculated spectra of Fig. 4b with a Gaussian of 46 meV FWHM (Fig. 4c), in order to take into account the energy spectral resolution determined by the experimental ZLP:

$$\text{EELS} = \text{ZLP} * \Gamma \qquad (4)$$

Fig. 4d shows the convolved calculated h-BN EEL spectra (shaded curves) for 2 nm, 15 nm and 30 nm thick h-BN slabs, respectively. They match remarkably well with the experimental h-BN EELS spectra (dots in Fig. 4d) obtained from the areas with the same h-BN flake thicknesses (marked by blue, red and green rectangles in Fig. 1a, respectively), thus corroborating our theoretical analysis and explanation of origin of the peak shifts.

**EELS near edges of vdW materials.** To understand the redshift of EELS spectra when the beam crosses the h-BN edge (see the area at the bottom flake edge in Fig. 1d), we numerically simulate (Comsol Multiphysics, see methods for details) the energy loss spectra for the electron passing directly at the edge (on-edge trajectory, see inset in Fig. 5e) of a 30 nm thick, semi-infinite slab (thick blue curve in Fig. 5a). Interestingly, we find two major differences with respect to the spectrum of an infinite h-BN slab (solid green curve in Fig. 5b): (i) the loss peak due to the guided wave excitation (marked by a red arrow at 174 meV in Fig. 5a) is redshifted by 2 meV and (ii) the bulk loss peak near the LO energy (marked by a green arrow in Fig. 5b) is shifted to 194 meV (1570 cm$^{-1}$, marked by a blue arrow in Fig. 5a). The lower energies of both peaks explain the redshift of the EELS spectra at the flake edge (Fig. 1d, bottom edge) however, do not provide the physical insights.

To understand the position of peaks in the EEL spectra at the flake edge, we calculated the cross sections of the electric field $E_{\mathrm{ind}}$ (evaluated at the energy loss $\Delta E$ =194 meV) induced by the electron beam for the on-edge and an the off-edge (away from the edge through the slab, see inset in Fig. 5f) electron trajectory (Fig. 5c and 5d, respectively). For the off-edge electron trajectory, we observe a zig-zag field pattern inside the h-BN slab, revealing the typical hyperbolic phonon polariton rays (see Fig. 3a). Above and below the h-BN slab (|z|>15nm), we find the field oscillations that correspond to the M0-HPhP mode[29,30,32]. For the on-edge electron trajectory, we additionally observe a strongly localized field at the h-BN edge, which leads us to assume that the electron beam excites a polariton mode propagating along the flake edge (along the y-axis). We support this assumption by analyzing the energy loss function $\Pi(q_y, \omega)$ (Fig. 5e), which was used to calculate the EEL probability $\Gamma(\omega) = \int \Pi(q_y, \omega) dq_y$ shown in Fig. 5a,b (see methods for details). $\Pi$ represents the probability of exciting a wave with momentum $q_y$ parallel to the h-BN edge (see inset in Fig. 5e) and is analogous to $P$ in equation (1). We see that for large momenta, $\Pi(q_y, \omega)$ asymptotically approaches the energy of 195 meV (vertical blue dashed line in Fig. 5e), in sharp contrast to $P(q, \omega)$ (Fig. 5f) that tends toward the LO energy. Interestingly, this asymptote at 195 meV corresponds to the surface optical phonon (SO) energy defined by the condition $\epsilon_\perp(\omega_{\mathrm{SO}}) = -1$ and is a signature of a surface polariton mode. We can thus conclude that the electron beam excites a hyperbolic *surface* phonon polariton (HSPhP) propagating at the edge surface of the flake.[11,12,33] . The peak in $\Gamma(\omega)$ at SO energy (blue spectrum in Fig. 5a) can thus be regarded as the loss due to surface phonon excitation[7]. We note, that surface polaritons do not exist on interfaces parallel to atomic layers in vdW materials and are normally disregarded. However, they can exist at the flake edges, as their surfaces exhibits the optical anisotropy (due to the atomic layers being perpendicular to the edge surfaces) required for surface polariton propagation[34,35].

The EEL peak at 174 meV (red arrow in Fig. 5a) can be also attributed to the excitation of the HSPhP at the flake edge. The spectral position of this peak, $\omega_{g,\mathrm{HSPhP}}$, is determined by the intersection of the "electron line" with the HSPhP dispersion, analogously to the energy loss due to guided wave excitation in the laterally infinite h-BN slab (see Fig. 3d,f). As the HSPhP dispersion has a steeper rise with energy compared to the HPhP dispersion, the peak position $\omega_{g,\mathrm{HSPhP}}$ = 174 meV is red shifted compared to the HPhP peak at $\omega_{g,\mathrm{HPhP}}$= 176 meV.

To obtain the complete picture of EELS at both sides of the flake edge, we numerically simulate EEL spectra $\Gamma(\mathrm{x}, \omega)$ for beam positions x on and outside the 30 nm thick, semi-infinite h-BN slab (Fig. 6a, top panel). For the electron passing through the h-BN flake (x<0), these spectra exhibit a small bulk loss peak at the LO energy (marked by the green arrow), which is consistent with our previous theoretical analysis (see green and dashed green curve in Fig. 3f and Fig. 5b, respectively). However, instead of a single peak due to the excitation of the M0-HPhP mode (red and dashed green curves in Fig. 3f and Fig. 5b, respectively), a set of fringes appears, whose position depends on the distance to the edge (marked by red arrows in Fig. 6a). These fringes can be understood as an interference pattern caused by the electron beam-launched M0-HPhP mode, which is reflected at the edge of the h-BN flake[29,30,36] (illustrated by the bottom schematics in Fig. 6a). This reflection can be phenomenologically introduced into our theory by adding a $\cos(2q|x| + \phi_{\text{refl}})$ in equation (1):

$$\Gamma(x,\omega) = \int_0^{q_c} P(q,\omega)[1 + \cos(2q|x| + \phi_{\text{refl}})]\, dq \qquad (5)$$

where $2q|x|$ is the propagation phase acquired by an HPhP after propagating toward the edge and back[36]; $\phi_{\text{refl}}$ is the phase acquired upon reflection[37]. We calculated $\Gamma(x,\omega)$ according to equation (5) and plotted its maxima as black dashed traces in Fig. 6a. We find a great agreement of these analytical lines with the fringe maxima obtained in Comsol simulations (color plot in Fig. 6a), proving that these fringes are due to HPhP interference and demonstrating the polaritonic nature of EEL in vdW materials.

For aloof beam trajectories (x>0 in Fig. 6a), neither the bulk loss peak near LO nor the surface loss peak near SO energy appear (marked by green and blue arrows, respectively) in the simulated EEL spectra. These peaks originate from the high-momentum excitations in the sample (see discussion below equation (2) and Supplementary Information 4), which are highly localized and therefore are inaccessible by aloof beams. As the result, only the peak associated with the excitation of guided waves is present in the aloof beam spectra. Remarkably, this peak matches the one in the on-edge spectrum (see the thin purple spectrum in Fig. 5a for better comparison), signifying that near the edges of vdW materials the electron predominantly excites the hyperbolic surface polariton modes and not the volume-propagating HPs.

To verify our analysis of the electron energy loss near the edges of vdW materials, we convolve the simulated spectra $\Gamma(x,\omega)$ in Fig. 6a with a 46 meV Gaussian function (to mimic the experimental spectral resolution) and compare the results (Fig. 6b) with the experimental EELS spectra (Fig. 6c) of h-BN collected along the orange line in Fig. 1a (see Methods for details). We see that the peak position $\omega_0$ of the convolved spectra (red curve in Fig. 6b) is in excellent agreement with that in our experiment (red curve in Fig. 6c), which supports the validity of our theoretical understanding. We also see that the current spectral resolution of our STEM-EELS is not sufficient yet for the direct observation of the interference fringes for x<0 (see Fig. 6a), which can be resolved by near-field optical microscopy[29,30,36]. However, we expect that the mapping of the HP interferences fringes becomes possible in the near-future with specialized low-energy STEM-EELS instruments that reach an energy resolution below 10 meV[14,38].

Interestingly, we see that the position of the h-BN peak is not constant for aloof beam spectra (x>0 in Fig.6b,c), but gradually redshifts with the distance to the h-BN edge. This is because the spatial probing range of an aloof electron (which can be regarded as an evanescent source of supercontinuum light[20]) decays exponentially with energy[39], which (for a given distance to the flake) suppresses the energy loss at high energies. The decay length is determined by $2\omega|x|/v$, and can extend tens to hundredths of nanometers (at mid-IR frequencies), which explains the nonvanishing energy loss outside the flake[14,40,41].

## DISCUSSION

We enabled a standard STEM-EELS system to detect low-energy excitations in vdW materials and used it to spatially map (for the first time with STEM-EELS) phononic excitations of h-BN at mid-IR. We demonstrated that the electron beam primarily probes not the bulk phonons, but the hyperbolic phonon polaritons, responsible for the unusual positions of the spectral peaks. We further showed that near the sample edges the electron primarily excites a hyperbolic surface polariton propagating along the edge surface of the vdW material – the finding important for the interpretation of spectra in aloof spectroscopy employed for avoiding the beam damage to fragile layered structures[42,43]. Our findings are supported by the dielectric theoretical framework for EELS of strongly anisotropic materials that could be applied to studies of phononic, plasmonic and excitonic media.

The dominance of hyperbolic polariton in EELS spectra of vdW materials makes STEM-EELS potentially highly suitable for the investigation of optical/polaritonic properties of such materials, with our work serving as a foundation for such investigations. With ongoing developments in monochromator and electron gun designs[38] – already ZLP as narrow as 10 meV in state-of-the-art specialized instruments[14] - we expect a significant improvement of the STEM-EELS spectral resolution in near future, allowing for correlative studies of polaritons and structural properties in vdW materials and turning the technique into a powerful complement to optical near-field methods.

## METHODS

**Sample fabrication.** We first performed mechanical exfoliation of commercially available h-BN crystals (HQ graphene Co., N2A1) using blue Nitto tape (Nitto Denko Co., SPV 224P). Then, we performed a second exfoliation of the h-BN flakes from the tape onto a transparent Polydimethyl-siloxane (PDMS) stamp. After that, via optical inspection of the stamp, a flake with desired topography was identified and transferred onto a suspended 15nm thick $Si_3N_4$ membrane (ultrathin TEM grid) using the dry transfer technique[44,45].

**Enhancing the primary beam monochromaticity and the spectral resolution of the detector in STEM-EELS.** EELS measurements were performed as STEM spectrum imaging at 60kV with a low-base Titan 60-300 TEM (FEI Co, Netherlands) equipped with a high brightness electron gun (x-FEG), electron beam monochromator and Gatan Quantum 965 imaging filter (GIF). The overall performance of (S)TEM-EELS system is mainly determined by two factors: the energy spread of electrons in the primary electron beam and the energy resolution of electron energy detection system, i.e. GIF.

To improve the monochromaticity (i.e. energy spread) of the primary electron beam we decreased the potential of the monochromator (operated with a standard excitation of 1-1.2) from a typical 3000V down to 700V while keeping the gun lens potential at around standard 700V. This way we moved the first gun crossover down from the accelerator to the plane close to the energy selecting slit. As a result, the crossover occurs for the electrons with a maximum speed for a given high tension. This dramatically decreases the energy spread broadening due to electron Coulomb interaction and therefore improves the electron beam monochromaticity.

To improve the energy resolution of EELS detection system we reduced the ZLP broadening due to finite size of energy GIF channel (usually estimated as 3 times the energy of a single channel) by implementing a special dispersion of 1.7 meV/channel (as measured by electrostatic drift tube of the GIF). Such dispersion is more than 5 times better than the smallest dispersion available at majority of commercial GIFs, reducing the ZLP broadening in the GIF by at least 25 meV. To further minimize the effect of aberrations (that negatively affects the EELS detection system resolution) we operated our microscope in microprobe STEM mode with a probe semi-convergence angle of about 1 mrad and extremely low camera length to obtain the collection angle of about 80 mrad.

The combination of system modifications described above, allowed us to reach on a regular basis an energy resolution of about 45-50meV per 1 msec of EELS acquisition time, and 38meV per 1 msec at peak performance. These results are rather reproducible and agree well with theoretical predictions[46]. Testing of our approach on another microscope (high-base Titan, 966 GIF) showed the same energy resolution also at 80 keV.

**Spectra acquisition, noise reduction and filtering.** Using the modified STEM-EELS system (see above) we have acquired a hyperspectral image of our sample which contains an EELS spectrum (with 1 msec acquisition time per spectrum) at each spatial pixel spaced 9.4 nm apart in both directions. Part of the image (about 1/3 of the total acquisition area directly above the area shown in Fig. 1a,d) was obtained with the electron beam switched off. The EELS signal from this area contains only the dark noise of the spectrometer. We averaged this dark noise among all spatial pixels (separately for each energy channel) and then subtracted it from every measured spectrum. Importantly this subtraction was performed channel-by-channel and prior to the energy axis calibration (i.e. assigning zero energy to the channel with maximum signal count), because the dark noise is channel specific.

The distribution of all spatial pixels was then evaluated based on the FWHM of the corresponding ZLP, which shows additional (to the main width distribution) peaks. These peaks could be attributed to mechanical vibrations and electromagnetic interference. We thus discard all pixels with ZLP width outside the main distribution and replace their spectra by interpolation from the nearest neighbors.

**Processing and analysis of experimental EELS spectra.** After the dark noise subtraction and pixels filtering (as described above), each EEL spectrum was normalized to its maximum. The signal from the bare membrane which we regard as ZLP, was then averaged in the region marked by magenta line in Fig. 1a to suppress the noise. This averaged ZLP was then subtracted from each spectrum yielding the h-BN inelastic scattering signal (i.e. energy loss to h-BN crystal lattice vibrations). To avoid possible drifts along the slow scan direction, the ZLP averaging and subtraction was performed separately for each horizontal line (that are aligned with the fast scan axis here). The h-BN inelastic scattering signal in the range from 140 to 220 meV was then fitted with a Gaussian, returning the best fit position of the peak maximum $\omega_0$. Finally, all the points with signal level below $2.5 \times 10^{-4}$ threshold or with irregular fits ($\omega_0$ outside of the upper Reststrahlen band or with FWHM<40 meV, i.e. below system spectral resolution) were filtered out and depicted in black in the map presented in Fig. 1d.

**h-BN film thickness estimation.** After performing EELS, the h-BN sample was scanned using atomic force microscopy (AFM) yielding a topography image which was then correlated with STEM-EELS image. The film thickness along the flake edges was determined through the step height measured as the AFM tip crosses from the membrane to the h-BN. To estimate the film thickness away from the edges, we associate the high-angle annular dark-field (HAADF) image contrast with the thickness, which can be done due to their monotonous interdependence. We then identify plateaus of the same contrast (marked by dashed white lines in Fig. 1a) and read off their thicknesses from the height at the plateaus' edges. This procedure yields the thicknesses of about 2, 15 and 30 nm for the blue, red and green areas marked in Fig. 1a, respectively.

**Line profile of the h-BN EELS spectrum across the flake edge.** The peak position in Fig. 6c was obtained by taking the h-BN spectra along the orange line marked in Fig. 1a. For each vertical position, spectra were averaged horizontally (7 pixels on both sides) and fitted with a Gaussian (curves in Fig. 6c), yielding the depicted profile of peak position, $\omega_0$.

**Full-wave numerical simulations of EELS probability.** The flake edge breaks the cylindrical symmetry, which prohibits the direct calculation of EEL according to equation (1). Therefore, EEL spectra near the h-BN edge (shown in Figs. 5 and 6) were simulated using finite element methods (Comsol Multiphysics) to solve the Maxwell's equations in the frequency domain. The moving electron (represented by a linear current) acts as a source, creating the electromagnetic field in the system. The calculation for each frequency is performed twice: with and without the dielectric environment (sample), preserving the same mesh. Their difference yields $\mathbf{E}_{\text{ind}}$, which can be integrated along the electron trajectory to find $\Gamma(\omega)$[47,48]:

$$\Gamma(x,\omega) = \frac{e}{\pi\hbar\omega}\text{Re}\left[\int_{-\infty}^{\infty} dz\, \hat{z} \cdot \mathbf{E}_{\text{ind}}(x,y,z;\omega)\exp\left(-i\frac{\omega z}{v}\right)\right],$$

where $x$ and y describe the lateral position of the electron across and along the edge, respectively (note that the energy loss is $y$-independent due to translational invariance of the system along the edge). Instead of directly solving for the induced field, we rewrite $\Gamma(x,\omega)$ using the Fourier transform (with respect to $y$) of $\mathbf{E}_{\text{ind}}(x,y,z;\omega) = \int_0^{\infty} dq_y\, \tilde{\mathbf{E}}_{\text{ind}}(x,z;q_y,\omega)\exp(iq_y y)$. By setting $y = 0$ (without the loss of generality) we can write:

$$\Gamma(x,\omega) = \int_0^{\infty} dq_y\, \Pi(q_y,\omega),$$

where $\Pi(q_y,\omega) = \frac{2e}{\pi^2\hbar\omega}\text{Re}\int_{-\infty}^{\infty} dz\, \hat{z}\cdot\tilde{\mathbf{E}}_{\text{ind}}(x,z;q_y,\omega)\exp\left(-i\frac{\omega z}{v}\right)$. This allows for replacing a single problem of finding $\mathbf{E}_{\text{ind}}$ in three dimensions with a set of 2D problems of finding $\tilde{\mathbf{E}}_{\text{ind}}$ as a function of x and $z$ for fixed $q_y$. We thus performed the calculations for discrete $q_y$ values up to the cutoff momentum $q_c$ and added up the individual contributions to find the total probability. Doing so, allowed for maintaining the same $q_c$ for different beam positions $x$, resulting in consistent calculation of the bulk loss. In addition, it allowed for the direct visualization of $\Pi(q_y,\omega)$ in Fig. 5e.

**ACKNOWLEDGEMENTS**

We thank the financial support of the European Union through ERC starting grants (SPINTROS grant no. 257654), the European Commission under the Graphene Flagship (GrapheneCore1, grant no. 696656), and the Spanish Ministry of Economy and Competitiveness (MAT2014-53432-C5-4-R, MAT2015-65159-R). We are also greatly thankful to S. Mastel for performing the AFM scan of the h-BN flake used to measure the flake thickness.

**ADDITIONAL INFORMATION**

**Competing financial interests:** The authors declare no competing financial interests.

**Authors contribution:** R.H. and A.C. conceived this study. S.V. and I.D. fabricated the sample and performed its optical characterization. F.C. and L.H. coordinated the sample fabrication. S.L. and A.C. performed the instrumental modifications of the STEM-EELS system. A.C. performed STEM-EELS and data processing. A.A.G. analyzed the data. A.K. and A.A.G. developed the analytical theory. J.A. supervised the theory development. A.K. performed the analytical and numerical calculations. A.Y.N. developed the framework for numerical calculations. R.H. coordinated and supervised the study. A.A.G. and R.H. wrote this paper with the input from A.K. All authors contributed to scientific discussions and manuscript revisions.

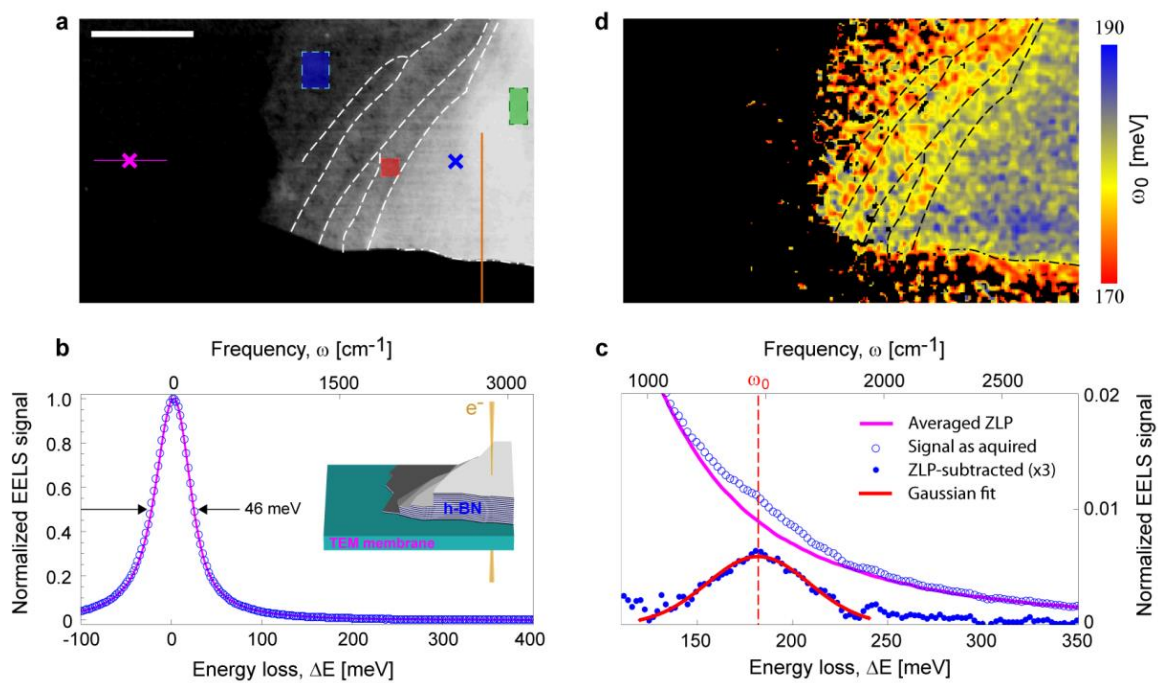

**Figure 1. EELS map of the hexagonal Boron Nitride flake. a,** STEM HAADF image: brighter areas correspond to larger h-BN thickness (see sample topology sketched in the inset of **b**). Scale bar is 500 nm. White dashed and dash-dotted lines are guides to an eye that mark pronounced steps in thickness and the bottom edge, respectively. Blue, Red and Green rectangles mark the area from which the spectra in Fig. 4d are collected. Orange line shows where the data for Fig. 6c is taken from. **b,** Typical spectra acquired on h-BN (Open Circles) and on an empty TEM membrane (magenta); the corresponding locations are marked, respectively, by magenta and blue crosses in **a**. Each spectrum is normalized to its ZLP maximum. **c,** Close-up view of the same spectrum as in **b** (open circles); the thick magenta curve shows the ZLP averaged along the horizontal magenta line in **a**. Closed circles and the red line show the h-BN vibrational signal after ZLP subtraction and the Gaussian fit to it, respectively, both magnified by a factor of 3 for better visibility. **d,** Map of the peak position $\omega_0$ that corresponds to the vibrational feature of h-BN. Black color marks regions of insufficient signal level (below $2.5 \times 10^{-4}$ threshold) and of otherwise irregular fits (see Methods for details). Black dashed and dash-dotted lines are the same eye guides as marked in **a**.

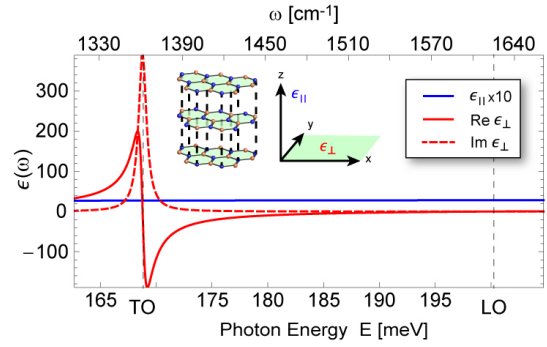

**Figure 2. Dielectric permittivity tensor of h-BN.** Real (solid red) and imaginary (dashed red) parts of the principal component of $\hat{\epsilon}$ perpendicular to the h-BN optical axis. For the parallel component the imaginary part is vanishingly small and only real part is shown (blue). Inset illustrates the orientation of tensor principal axes with respect to the atomic sheets of h-BN. Left and right vertical dashed line mark the TO and LO phonon frequencies, respectively.

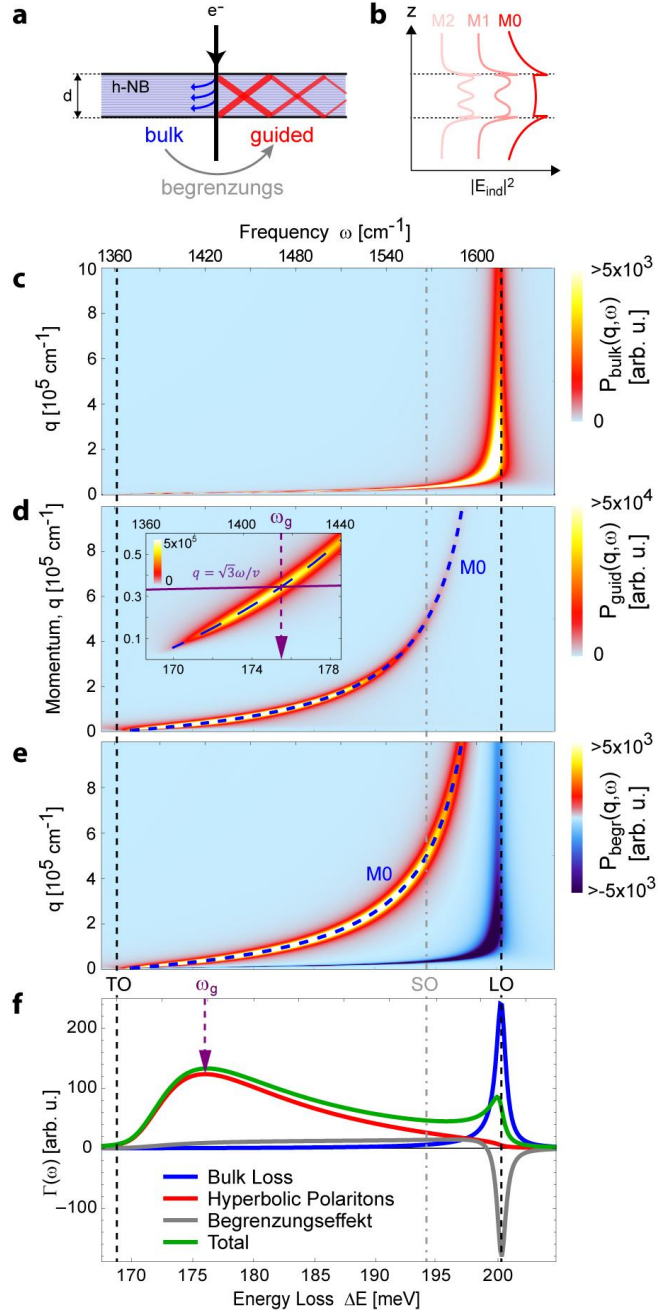

**Figure 3. Theoretical EEL spectrum of a homogeneous h-BN slab. a,** A schematic representation of electron energy loss due to excitation of bulk phonon (left) and guided modes (right) in a dielectric slab. **b,** Intensity profiles of three guided HPhP modes. The energy-momentum maps of $P_{bulk}(q,\omega)$ **(c)**, $P_{guid}(q,\omega)$ **(d)** and $P_{begr}(q,\omega)$ **(e)** calculated for a 30 nm thick h-BN slab. The dashed curves in **d** and **e** depict the dispersion of M0 mode of the hyperbolic phonon polariton (calculated according to equation S27)[30]. Inset in **d** is a close up on $P_{guid}(q,\omega)$ at low momenta; dashed purple arrow shows the energy at which the "electron line" $q = \sqrt{3}\omega/v$ (solid purple line, corresponding to the maximum of momentum transfer from electron to the guided mode) intersects with the M0-HPhP mode dispersion. **f,** EEL probabilities, $\Gamma(\omega)$, corresponding to bulk (blue), guided-wave (red), begrenzungseffekt (gray) losses and their sum (green). The vertical purple arrow marks the same energy as in inset

to **d**. In all plots the vertical dashed lines mark the TO and LO phonon frequencies; vertical dot-dashed line marks the location of the surface polariton frequency (corresponds to $\epsilon_\perp = -1$ here).

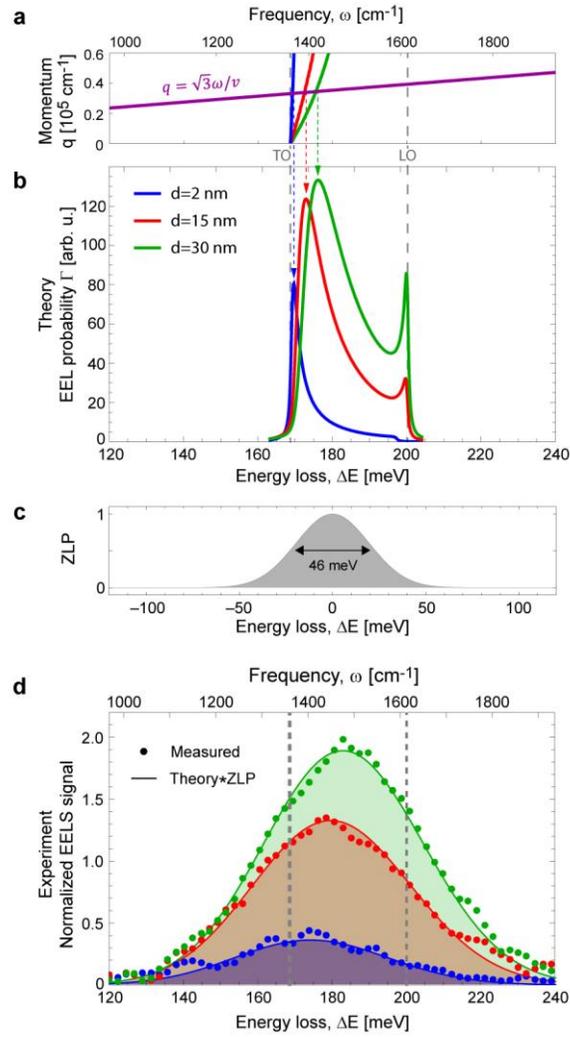

**Figure 4. Thickness dependence of h-BN EEL spectra. a,** Dispersions (low energy part) of the fundamental M0-HPhP mode in h-BN films of 2 nm (blue), 15 nm (red) and 30 nm (green) thicknesses. Blue, red and green vertical dashed lines mark the energies at which these dispersion curves intersect the "electron line" $q = \sqrt{3}\omega/v$ (purple line), and which determine the positions of spectral maxima of $\Gamma_{\text{guid}}(\omega)$. **b,** Theoretically calculated EELS probability for the corresponding film thicknesses. **c,** A Gaussian with FWHM = 46 meV representing the experimental ZLP. **d,** Experimental spectra (dots) of h-BN averaged over areas marked by blue, red and green boxes in Fig. 1a. The shaded curves are theoretical EEL spectra obtained by convolving spectra of **b** with the ZLP in **c**. The magnitude of all calculated spectra was scaled by the same factor to correspond with the experiment. In all plots the left and right vertical dashed lines mark the TO and LO phonon energies, respectively.

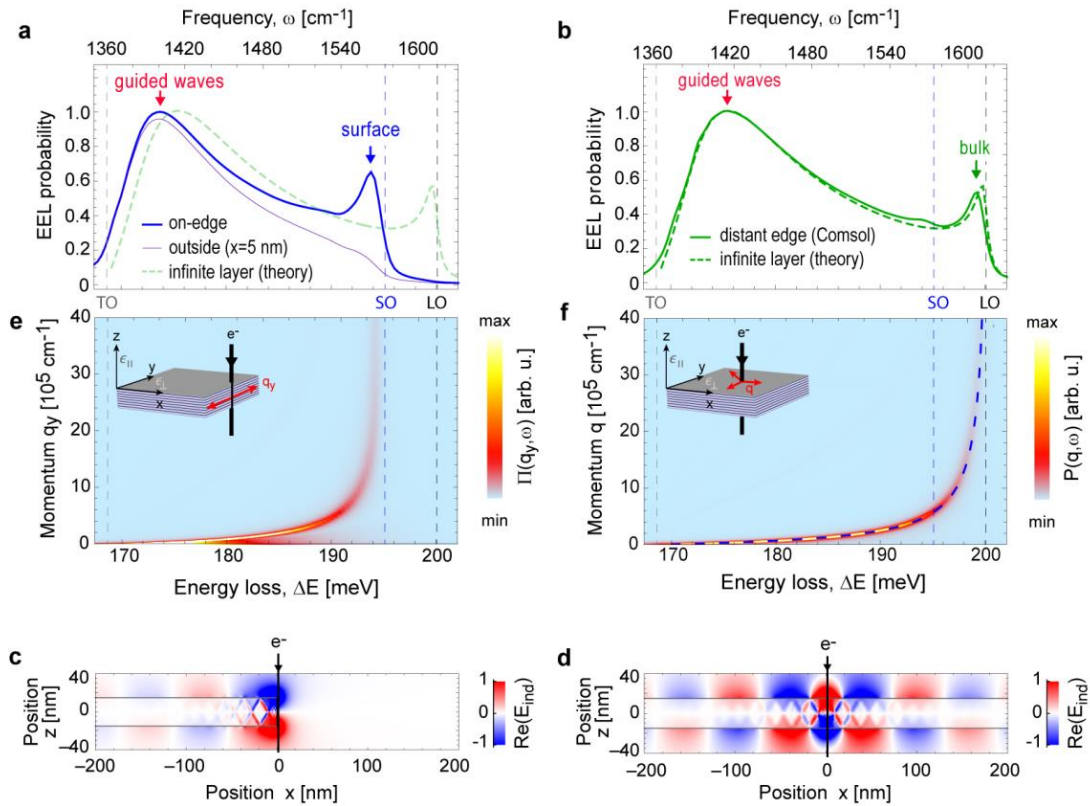

**Figure 5. Volume vs. edge-localized hyperbolic polaritons. a,** EELS spectrum of 30 nm thick semi-infinite slab for on-edge electron trajectory (blue) and 5 nm outside (thin purple) obtained using full-wave simulation (Comsol). **b,** EELS spectrum for the beam passing through the h-BN slab away from the edge (off-edge trajectory) obtained using Comsol (solid green). Dashed green spectra in **a** and **b** are retarded theoretical calculations (see Supplementary Information 5) using equation (1). **c,** Cross section of the induced electric fields (z-component) taken at $\omega = 1570$ cm$^{-1}$ (marked by blue arrow in **a**) for the on-edge electron trajectory. **d,** Same as in **c** but for off-edge electron trajectory. **e,** Momentum dependent probability $\Pi(q_y, \omega)$ for on-edge eletrcon trajectory. **f,** $P(q, \omega)$ for a 30 nm thick, laterally-infinite layer of h-BN (the sum of contributions depicted in Fig. 2a,b, and c with retardation included, see Supplementary Information 5). Insets in **e** and **f** show the electron trajectories and the polariton propagation directions. In all plots the vertical dashed lines mark the TO, SO and LO phonon energies.

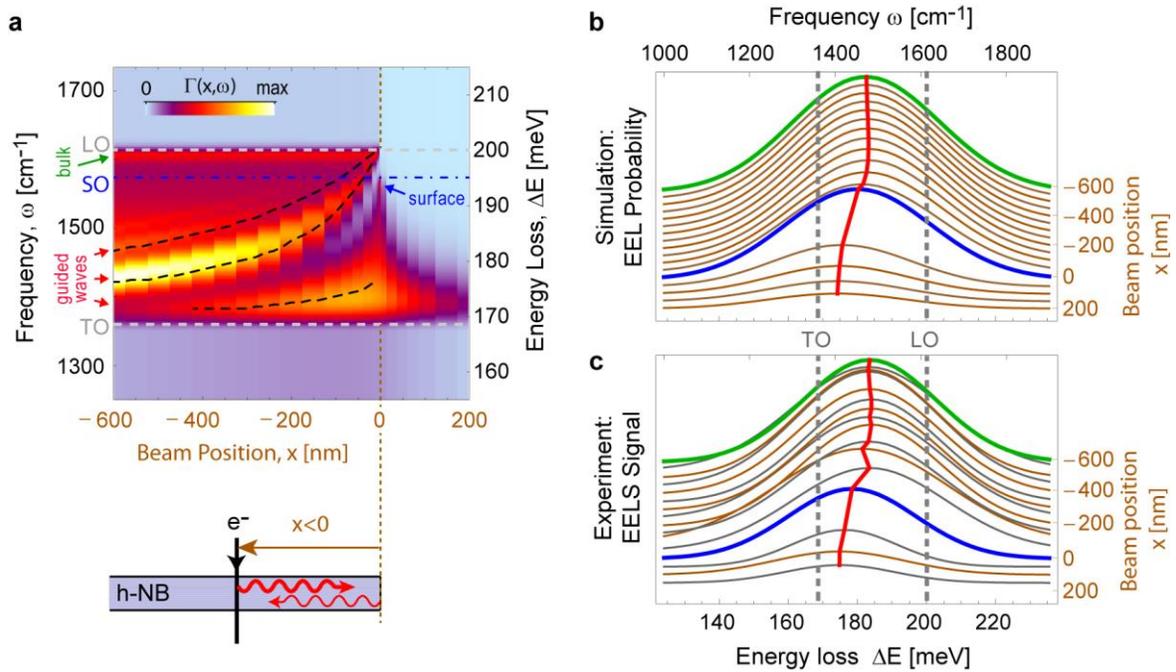

**Figure 6. EELS spectra near film edges. a,** Numerically calculated (Comsol) EELS probability spectra for 30 nm thick semi-infinite h-BN slab as a function of distance to its edge. Black dashed curves trace the interference fringe maxima appearing due to polariton reflection from the film edge (schematically shown at the bottom) calculated using equation (5) with reflection phase $\phi_{\text{refl}} = \pi/2$. **b,** Same as in **a**, but after convolution according to equation (4). The spectra are offset vertically for better visibility. Red line marks the position of the peak maximum. Blue spectrum corresponds to the convolution of the solid blue spectrum in Fig. 5a. The vertical dashed lines mark the TO and LO phonon energies. **c,** Gaussian fits to the experimental EELS spectra for different beam positions near the edge of the 30 nm thick part h-BN flake. The spectra are collected along the orange line marked in Fig. 1a (see Methods). Red line marks the position of the peak maximum, $\omega_0$.